\documentclass[%
 reprint,
superscriptaddress,
 amsmath,amssymb,
 aps,
]{revtex4-2}
\usepackage[dvipsnames]{xcolor}
\usepackage{graphicx}
\usepackage{dcolumn}
\usepackage{bm}
\usepackage{hyperref}
\newcommand{\ket}[1]{|#1\rangle}

\begin{document}

\preprint{APS/123-QED}

\title{On readout and initialisation fidelity by finite demolition single shot readout}

\author{Majid Zahedian}
 \affiliation{3rd physical institute University of Stuttgart, Germany}
  \affiliation{Friedrich-Alexander-Universit\"at Erlangen-N\"urnberg, Germany}

\author{Max Keller}
 \affiliation{3rd physical institute University of Stuttgart, Germany}

\author{Minsik Kwon}
 \affiliation{3rd physical institute University of Stuttgart, Germany}
  \affiliation{Max Planck Institute for solid state physics, Stuttgart, Germany}
 
 \author{Javid Javadzade}
 \affiliation{3rd physical institute University of Stuttgart, Germany}

\author{Jonas Meinel}
 \affiliation{3rd physical institute University of Stuttgart, Germany}
  \affiliation{Max Planck Institute for solid state physics, Stuttgart, Germany}

\author{Vadim Vorobyov}
\email[]{v.vorobyov@pi3.uni-stuttgart.de}
 \affiliation{3rd physical institute University of Stuttgart, Germany}  

\author{J\"org Wrachtrup}
 \affiliation{3rd physical institute University of Stuttgart, Germany}
 \affiliation{Max Planck Institute for solid state physics, Stuttgart, Germany}

\date{\today}

\begin{abstract}
Ideal projective quantum measurement makes the system state collapse in one of the observable operator eigenstates $|\phi_\alpha\rangle$, making it a powerful tool for preparing the system in the desired pure state. 
Nevertheless, experimental realisations of projective measurement are not ideal.
During the measurement time needed to overcome the classical noise of the apparatus, the system state is often (slightly) perturbed, which compromises the fidelity of initialisation.
In this paper, we propose an analytical model to analyse the initialisation fidelity of the system performed by the single-shot readout. 
We derive a method to optimise parameters for the three most used cases of photon counting based readouts for NV colour centre in diamond, charge state, nuclear spin and low temperature electron spin readout. 
Our work is of relevance for the accurate description of initialisation fidelity of the quantum bit when the single-shot readout is used for initialisation via post-selection or real-time control. 
\end{abstract}

\maketitle
\section{Introduction and background}
Quantum projective measurement of a single system is intrinsically probabilistic, and the probability of measuring an eigenvalue $\alpha$ reads $p_\alpha = \mathrm{Tr}(P_\alpha\rho)$.
Once $\alpha$ was measured, the system collapses ($reduces$) to the state $\rho_\alpha = P_\alpha \rho P_\alpha^\dag /p_\alpha$. 
Therefore, successive projective measurements of observable $\mathbb{O}$ have probability $p_\alpha'= \mathrm{Tr}(P_\alpha P_\alpha \rho P_\alpha^\dag/p_\alpha) = 1$ to give the same output $\alpha$ and hence same post-measurement state.  
Due to the preservation of the once measured ($collapsed$) state, its experimental realisation is referred to as quantum non-demolition (QND) measurement, which was demonstrated in many systems, e.g., trapped ion, dopants in diamond, superconducting qubits \cite{Wineland_SSR, NeumannSSR, dreau2013single, robledo2011high, sukachev2017silicon, chen2020parallel, Shields2015, zhang2021high,Irber_2021, Makhlin2001uo}. 
In many cases, measurements of the quantum system are associated with obtaining a macroscopically readable signal, usually is encoded in the number of photons or electrons. 
Thus, the counting process is stochastic and classical, meaning  it does not relate to the $quantumness$ of the measured system. 
This process determines the measurement noise. 
In most cases, this process could be described as a Poissonian distribution with a mean number of counts $\bar{\lambda}$. 
Various states of the system generate distributions of meter outputs with various average values $\bar{\lambda}_i$. 
Due to the finite width of the distributions, there is always an overlap between them. 
This overlap causes uncertainty in the state estimation \cite{kay1993fundamentals}. 
With a cycling non-demolition measurement, the signal could be acquired as long as needed to achieve the desired fidelity.
In particular, when the signal-to-noise ratio acquired within the measurement time of the system is above unity, it is denoted as $single \, shot \, readout$ corresponding to fidelity exceeding 79 \% \cite{Hopper2018}. 
In practice, most experiments subject the system to additional decay channels, limiting the available measurement time and , consequently, its fidelity. 
Also, it disturbs the state of the system during the measurement. 
So, it could scrutinise the state preparation method, which relies on the measurement, such as post-selection or active feedback drive for the desired state preparation. 
Therefore accurate estimation of the initialisation fidelity is essential for benchmarking and optimising the quantum hardware. 
It became crucial for estimating the feasibility and performance of envisioned quantum algorithms such as surface codes \cite{fowler2012surface}.
Although such fidelity could be optimised experimentally \cite{Shields2015}, a universal analytical solution is desirable for fast optimisation of readout parameters. 

When the system spontaneously changes its state during the measurement, the macroscopic outputs probability distributions ($PDF_{0,1}$) are not Poissonian anymore. 
Additionally, the system changes its states during the readout, and the moment $t$ at which the system was in $\ket{0}$ or $\ket{1}$ state shall be specified to calculate the $PDF_{0,1}^{t}$ accurately. 
One approach is to fix the time at the beginning of the measurement $t=0$ and derive the $PDF_{0,1}^{t=0}$. 
Having a meter reading $\lambda_i$ using the maximum likelihood method, one could estimate in which state system was at the moment $t=0$ \cite{Gambetta2007vx, DAnjou2014tz, Shields2015, Hopper2020}. 
Using techniques of statistical signal analysis \cite{kay1993fundamentals}, the fidelity of readout could be optimised by choosing the optimum detection threshold $\lambda_{th}$, as well as using filtering techniques \cite{Gambetta2007vx,DAnjou2014tz, DAnjou2016up,DAnjou2017us}. 
However, when estimating the state of the system at the end of the measurement $t=T$, one has to make additional assumptions about state decay during the measurement.
\begin{figure*}
\begin{center}
\includegraphics[width=0.9\textwidth]{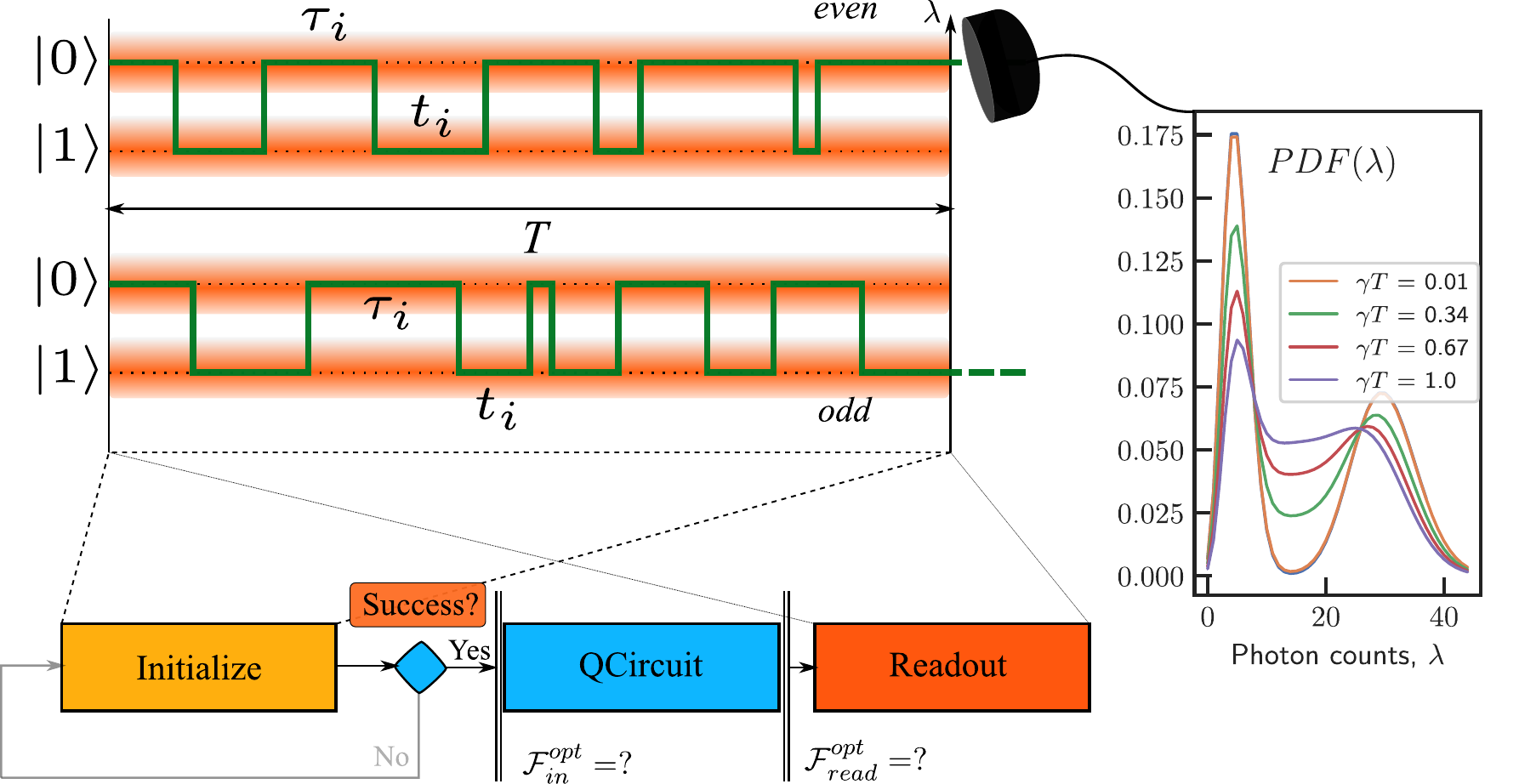}
\caption{Readout during the switching dynamics of the two-level system. The switching rate $\gamma$ and measurement time $T$ ratio determines the amount of overlapping of the measurement output $\lambda$ distributions for various states. Depending on the task, the parameters could be optimised in order to either initialize the system by measurement or readout the system state.}
\label{fig:1}
\end{center}
\end{figure*}
In this work, we follow an approach of directly using measurement statistics to infer the fidelity of the final state. 
By setting the aforementioned time when defining the PDFs to the end of the measurement $t=T$, we calculate the $PDF_{0,1}^{t=T}$ and directly apply the maximum likelihood method for the readout of the final state. 
This approach allows for more accurate fidelity estimation of state preparation using finite demolition readout in the presence of noise.
We apply this model to the NV centre low-temperature electron spin readout, room temperature charge state readout, and nuclear spin single shot readout and use experimental data to build the model and optimise the parameters. 


\section{Theory}
\label{sec:theory}

We recall the main expressions for dynamics of the two-state Markov process with exponentially distributed switching times. 
By using the properties of the point Poisson process (see Appendix \ref{app:theory}), we obtain the probability distribution of the time spent by the system in state $\ket{0}$ and $\ket{1}$ conditioned on the initial state of the system (eq. \ref{eq:prob_0} in Appendix \ref{app:theory}). 
The distribution of photon counts is obtained by assuming the emission of photons is a Poisson process, with an average per time being the average emission rate of the system in each state. 
The probability of emitting a certain amount of photons $\lambda$ is then calculated by integrating the probability to spent time $\tau$ in state $\ket{0}$ weighted by the Poissonian probability distribution to emit n photons with average emission rate $\lambda = \lambda_0 \tau - \lambda_1 (T-\tau)$.  
Finally, the distribution of photon counts conditional on the initial state of the system is equivalent to the one obtained in previous work {\cite{Shields2015}.
The overall photon counting statistics are numerically simulated and presented in figure \ref{fig:1}.
The distributions deviate from the Poisson distribution, and the overlap between the distributions increases with increasing switching rates.
In the case where $\gamma =  T^{-1}$ the distributions significantly overlap, hence the measurement has less information to resolve the state, and the fidelity of the measurement $\mathcal{F}$ reduces, while in the case where $\gamma \le T^{-1}$ despite a significant overlap between the distributions one could discriminate the initial states.

To infer the photon counting probability distribution conditioned on the final state after the measurement, we look into the switching dynamics using Bayes's conditional probability rule. 
Knowing the probability distributions of time spent in state $\ket{0}$ and $\ket{1}$ we obtain:
\begin{equation}
 p(\lambda | \ket{0}_T) = p(\lambda | (\ket{0}_0 \cap even) \cup (\ket{1}_0 \cap odd)) 
\end{equation}
And from the rules of conditional probability, it follows: 
\begin{widetext}
\begin{equation}
	\begin{split}
		p(\lambda | (\ket{0}_0 \cap \mathrm{even}) \cup (\ket{1}_0 \cap \mathrm{odd})) &= \frac{p(\lambda \cap \ket{0}_0 \cap \mathrm{even})+p(\lambda \cap \ket{1}_0 \cap \mathrm{odd})}{p(\ket{0}_0 \cap \mathrm{even})+p(\ket{1}_0 \cap \mathrm{odd})}\\
		=& \frac{p(\lambda \cap \mathrm{even} | \ket{0}_0 ) p(\ket{0}_0) + p(\lambda \cap \mathrm{odd} | \ket{1}_0) p(\ket{1}_0)} {p(\ket{0}_0)p(\mathrm{even}|\ket{0}_0 )+p(\ket{1}_0)p( \mathrm{odd}|\ket{1}_0)} \\
	\end{split}
	\label{eq:final_p}
\end{equation}
\end{widetext}
The probability of having initial state $p(\ket{0/1}_0)$ in the general case could be inferred by fitting photon counting statistics with model conditional on initial states \ref{eq:prob_n}. 
Probabilities in the case of the continuous measurements in the steady-state are $p(0/1) = \frac{\gamma_{1,2}}{\gamma_1 + \gamma_2}$.
$P(even | \ket{0})$ could be obtained by e.g. integrating  $ \int_0^T d\tau \, P(\tau \cap even | \ket{0})$. 
\begin{figure}
\begin{center}
\includegraphics[width=\columnwidth]{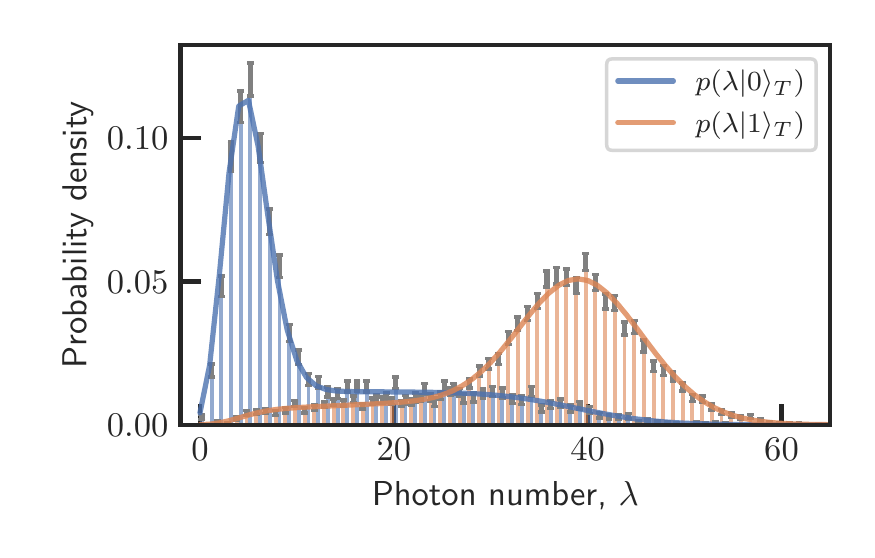}
\caption{Comparison of analytical and Monte-Carlo simulation of PDF conditional to final state. Parameters:  $\gamma_0 = 500$ Hz, $\gamma_1 = 300$ Hz, $\lambda_0 = 5$ kHz, $\lambda_1 = 40$ kHz, T= 1 ms, 10000 runs}
\label{fig:MC_final}
\end{center}
\end{figure}
We compare the probability distributions of photon counts conditioned on the final state estimated using analytical expression \ref{eq:final_p} and simulated numerically via the Monte Carlo method. For this case, we considered a probable scenario with 1 ms readout time and decay rates of 500 Hz and 300 Hz with photon counts of 5 kHz and 40 kHz. 
The shape of the distributions is reproduced by the numerical simulation, and analytical expression catches qualitatively and quantitatively the shape of the distribution.

Knowing the analytical expression for probability density function for detector output conditioned on the final state of the system, we derive a likelihood function $L(\theta | \lambda) = p(\lambda | \theta)$, where $\theta$  represents a discrete set of final states 0 and 1.
By introducing the $\Lambda = L(0| \lambda) / L (1| \lambda)$ one attributes the final state to 0 if $\Lambda > 1$ and to 1 if else, the state is attributed to randomly tossed coin result in case $\Lambda = 1$. 
We note that if the probability of having a certain final state depends on measurement parameters, the expressions should be weighted by the probability of a certain final state $\theta$: $L'(\theta | \lambda) = p(\theta) \cdot p(\lambda | \theta)$.
The error rate of such a decision is thus the number of cases where the decision was falsely made with such a strategy; hence it is the overlapping area under the probability distributions. 
This area could be numerically estimated and minimised with respect to the parameters of the readout, such as readout duration and, if possible, switching rates and photon emission rates, when using a known models for optical excitation. 
As an example, we estimate the fidelity numerically for the realistic case of single shot-readout of the nuclear spin at room temperature \cite{NeumannSSR} for the various nuclei with spin decay rates. The photon count rates of NV in solid immersion lens normalized to the average duty cycle of the excitation laser result in rates $\lambda_0 = 100$ kHz for state $\ket{0}$ and $\lambda_1= 70$ kHz for state $\ket{1}$. 
\begin{figure}
\begin{center}
\includegraphics[width=\columnwidth]{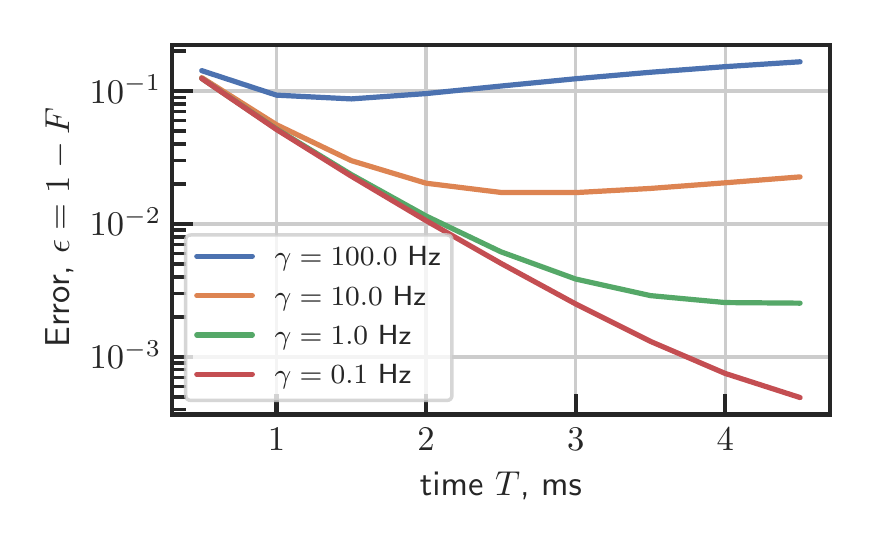}
\caption{The error rate for posterior state estimation without data loss. Parameters:  $\gamma_{0}=\gamma_{1} = 0.1-100$ Hz, $\lambda_0 = 70$ kHz, $\lambda_1 = 100$ kHz, T= 0.5-4.5 ms}
\label{fig:fid_simple}
\end{center}
\end{figure}
As seen in figure \ref{fig:fid_simple} the error rate reaches the minimum for realistic cases at $T= 2.5$ ms and $T=1.5$ ms accordingly for $\gamma = 10$ and $\gamma=100$ Hz.

To see the advantage of using such a method for estimating fidelity, we calculate the error rate of post-selection when using the decision-making condition $\lambda > n_{th}$. By increasing $n_{th}$, one can select part of the distribution corresponding to the bright state $\ket{1}$, such that the tail of dark state $\ket{0}$ is excluded. 
This method thus allows to increase the fidelity by sacrificing measurement efficiency.
The region of intersection is excluded from the decision, thus reducing the sample volume of the dataset. 
Although for the case of initialization by measurement, the initial state is known with high precision, to estimate the final state, one had to estimate the probability that the system stayed unperturbed.  
In a simple case, this could be done by multiplying the fidelity of estimating the initial state by the exponential decay, which approximates the probability of the system not relaxing.
As seen in figure \ref{fig:fid_exclusive}, this sets the lower estimate of fidelity for lower data usage. 
In practice, when using a posterior estimation, much higher fidelities could be reached, with error rates of several orders of magnitude less, although it could be achieved at low-efficiency rates. 
\begin{figure}[h]
\begin{center}
\includegraphics[width=\columnwidth]{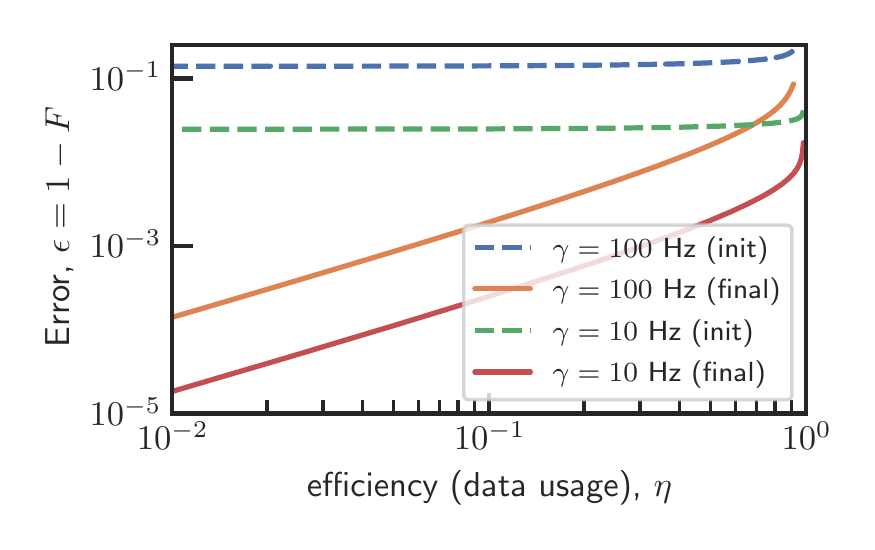}
\caption{The error of post-measurement state estimation versus the fraction of used data (efficiency). The dashed line is an estimation error of the initial state combined with the probability that the system stayed unperturbed during the readout. The solid line is the maximum likelihood method using pdf conditional to the final state. Parameters:  $\gamma_{0}=\gamma_{1} = 10,100$ Hz,  $\lambda_0 = 70$ kHz, $\lambda_1 = 100$ kHz, T= 1.5 and 2.5 ms respectively for 10 and 100 Hz}
\label{fig:fid_exclusive}
\end{center}
\end{figure}
\section{Results}
\label{sec:result}
We applied the developed theoretical framework for optimizing the experimental parameters of readout for initialization fidelity in three real scenarios of the well-studied model system of NV centre in diamond. 
We consider the case of room temperature charge state initialisation, room temperature single-shot readout of strongly coupled nuclear spin, and low-temperature resonant readout of the electron spin. 

\subsection{Electron spin readout at low temperature}
\begin{figure}[h]
	\begin{center}
		\includegraphics[width=\columnwidth]{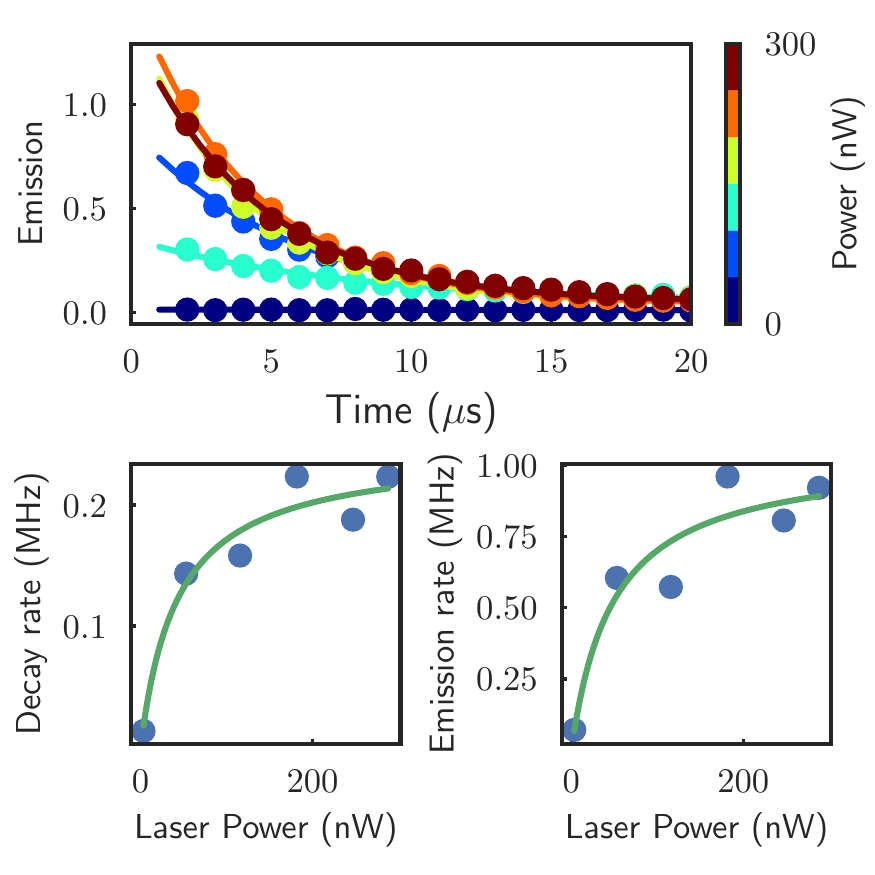}
		\caption{Experimental calibration of electron spin readout. Top plot - statistic of photons arrival time, during the measurement time t. Bottom left - decay rate of the $m_s = 0$ state $\gamma$ as a function of laser power. Bottom right panel: Emission rate of the photons of the bright state as a function of laser power.}
		\label{fig:electron_ssr_param}
	\end{center}
\end{figure}

\begin{figure}[h]
\begin{center}
\includegraphics[width=\columnwidth]{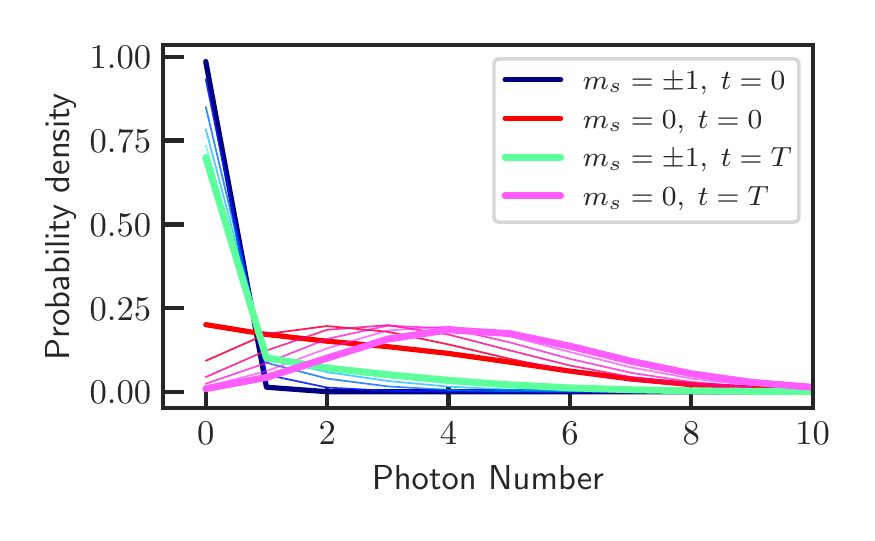}
\caption{The probability distributions conditional on the state of the system at the moment $t$ during the measurement of duration T. Blue curves conditional to $m_s \ne 0$, red curves $m_s = 0$. Cayenne and Pink curves are conditional to the state corresponding to the final moment T. Washed curves indicate transition of PDFs conditioned to start of the measurement $t=0$ to finish $t=T$}
\label{fig:fidelity_time}
\end{center}
\end{figure}

We start by considering the case of low-temperature resonant electron spin readout. 
In this case,  the switching rates follow $\gamma_0 \gg \gamma_1$, so we can neglect the $\gamma_1$.  
As a result, the distributions could be significantly simplified:
\begin{equation}
\begin{split}
&p(\lambda|\ket{0}_0) = \int_0^T dt e^{-\gamma t}\cdot \mathrm{Poiss}(\lambda, \lambda_0 t + \lambda_1 (T-t)) \\
&p(\lambda|\ket{1}_0) = \mathrm{Poiss}(\lambda, \lambda_1 T) \\
&p(\lambda|\ket{0}_T) = \mathrm{Poiss}(\lambda, \lambda_0 T) \\
&p(\lambda|\ket{1}_T) = p(\ket{1}_0) \cdot \mathrm{Poiss}(\lambda, \lambda_1 T)  + p(\ket{0}_0) \cdot  \\
&\int_0^T dt e^{-\gamma t}\cdot \mathrm{Poiss}(\lambda, \lambda_0 t + \lambda_1 (T-t)).
\end{split}
\end{equation}
\begin{figure}[ht]
\begin{center}
\includegraphics[width=0.5\textwidth]{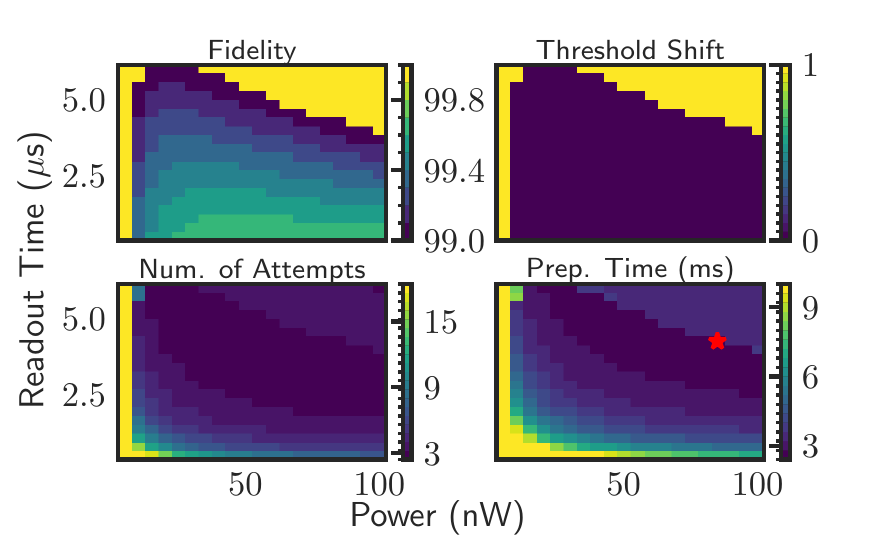}
\caption{Electron spin readout for state ms=0 a) fidelity, b) threshold shift, c) number of attempts, d) required time}
\label{fig:readout_ex}
\end{center}
\end{figure}

We first calibrate the optical parameters in the experimental setup and extract the decay rate and emission parameters of the NV system (figure \ref{fig:electron_ssr_param})
The numerical simulation of the distributions conditional on initial and final states for a set of the excitation laser and readout time is presented in figure \ref{fig:fidelity_time}.
This plot shows how conditional distributions transform with time at which the condition of certain state is taken from $t=0$ to $t=T$. 
At $t=0$, the distributions present a well-known shape \cite{robledo2011high}, which we observe in our experiments by preparing the initial state into $m_s =0$. 
The distribution transforms by moving the conditioning to the measurement's end.  
If the final state is $m_s = 0$, the distribution becomes purely Poissonian with average $\lambda = \lambda_0 T$ since no jump occurred. 
While if the final state is $m_s = \pm1$, the distribution is mixed and can be calculated as integral.   
To reach high fidelities, reading (selecting) only of the state $m_s=0$ is applied.
We simulate the fidelity based on the formula $\mathcal{F} = B/(B+D)$, where $B$ is the area under the distribution above the threshold of the bright state $m_s = 0$, and D is the area under the distribution of dark state $m_s =\pm1$. 
Depending on the readout power and duration, we find the necessary threshold that guarantees a target fidelity of 99\%. For the readout of state 0, we see that already threshold 0 is enough in most cases to achieve the desired fidelity. 
Next, we plot the average number of attempts, which is inversely proportional to the ratio between the selected area above the threshold and the overall area under the distributions. 
The average number of attempts determines the success rate of the readout. It is used to find the optimal readout parameters that minimize the time necessary to measure a single data point with a target fidelity. 
We find that it is $2.55$ ms for the optimal parameters laser intensity of $85$ nW and the readout time of 4.26 $\mu s$ assuming one sequence is 1 ms on average. 
\begin{figure}[h]
\begin{center}
\includegraphics[width=0.5\textwidth]{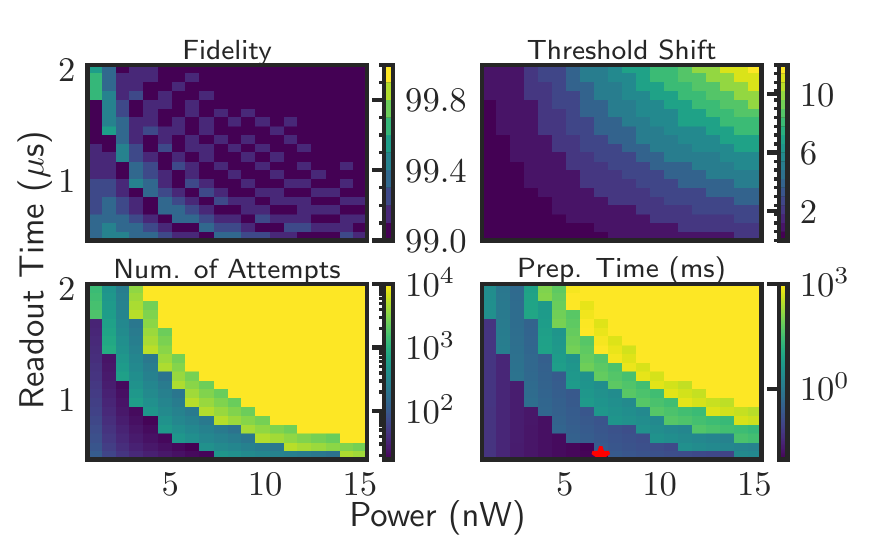}
\caption{ After 90\% preinitialisation, electron spin preparation into $m_s= 0$ a) fidelity, b) threshold shift, c) number of attempts, d) required time}
\label{fig:preparation_ex}
\end{center}
\end{figure}
Now we consider the case of preparing the desired state $m_s = 0$ by the measurement. 
In this case, we again visualise fidelity, but the drastic change is the necessary threshold, which is needed to be applied in order to achieve the desired fidelity. 
We note that the distributions conditional on the final state, in this case, are weighted by the probabilities of the final state, which tend to decay towards a steady state upon readout $p_0(t) = p_0 exp(-\gamma t)$, $p_1(t) = 1- p_0(t)$. 
In this case, the success rate of initialisation is significantly reduced, and the desired fidelity is achieved in 9.22 $\mu s$ with optimal parameters 6.89 nW and 0.5 $\mu s$, which differ from the case of readout. 

\subsection*{NV charge state readout at room temperature}

Next, we consider the case of the charge state readout of the NV centre at room temperature.
We use orange laser (594 nm) excitation and long pass 650 nm filter to exclude the NV0 fluorescence, achieving high contrast between states $NV^-$ and $NV^0$. We calibrate the fluorescence photon counting rate $\lambda_{0,1}$ and state switching rate $\gamma_{0,1}$ as a function of laser intensity. 

\begin{figure}[ht]
\begin{center}
\includegraphics[width=0.5\textwidth]{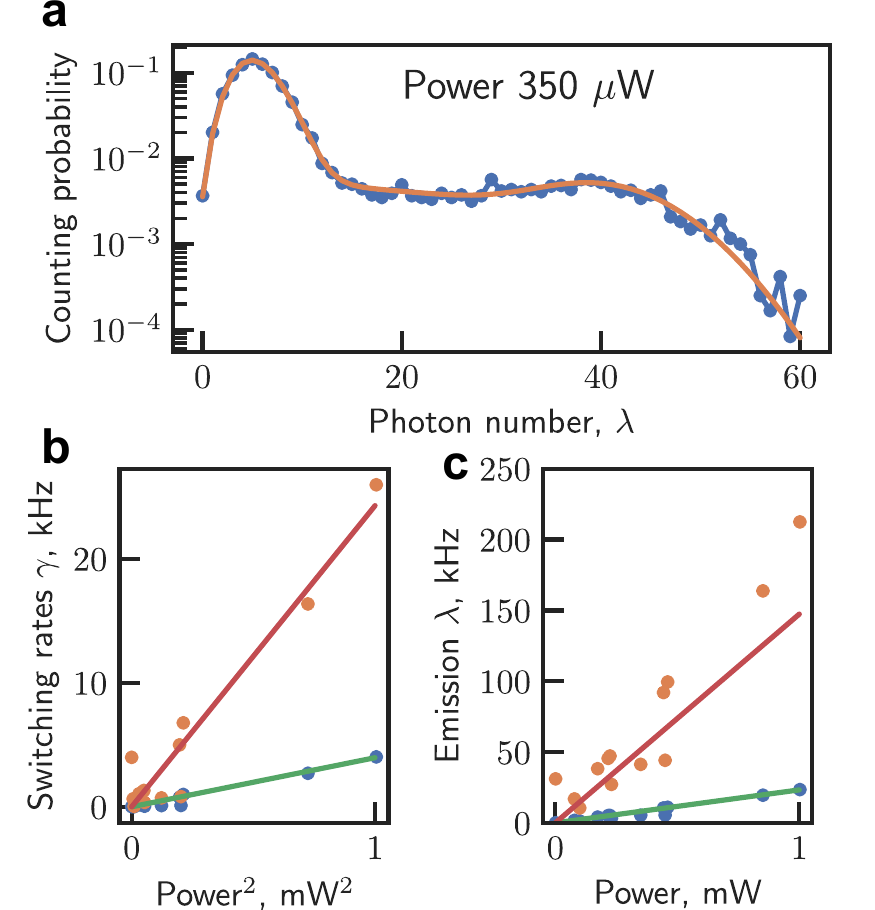}
\caption{Experimental charge state rates calibration a) Exemplary histogram of photon counts with laser intensity 350 $\mu$W b) The extracted switching rate in kHz between two states as a function of laser intensity c) the extracted fluorescence rate in kHz as a function of laser intensity.}
\label{fig:charge_state_calibration}
\end{center}
\end{figure}

\begin{figure}[ht]
\begin{center}
\includegraphics[width=0.5\textwidth]{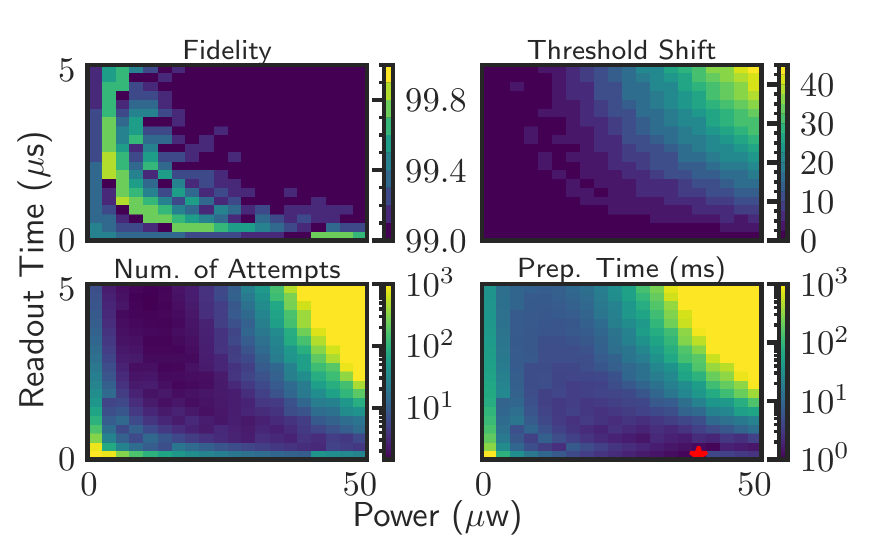}
\caption{Charge state preparation a) fidelity, b) threshold shift, c) number of attempts, d) required time}
\label{fig:charge_state_time}
\end{center}
\end{figure}
We consider the case of charge state $NV^-$ initialisation. 
It is commonly done by applying a short green laser pulse and a weak orange or red probe readout pulse.  
Depending on the photon counts during the orange probe, the state can be assigned to be in $NV^-$. 
A feedforward operation for on-demand state initialisation could be applied \cite{Hopper2020}. 
In this section, we optimise readout parameters concerning the preparation time of the charge state. 
We consider several target fidelities for preparation. In the main text, we present only the case of $\mathcal{F} = 99\%$ and other cases presented in SI.

Using the formulas and the model of the defect charge switching and photon emission, for each parameter of orange laser power and duration of the readout, we calculate the probability distribution function and estimate fidelity represented in figure \ref{fig:charge_state_time}a. 
To reach a target fidelity, we apply the exclusion principle and increase the photon number threshold, thus reducing the efficiency of the readout, which leads to an increase in the number of attempts of a successful measurement. 
Accordingly, we plot a required increase in the threshold in figure \ref{fig:charge_state_time}b. 
Then using the success rate of a single measurement, we estimate the average number of attempts (figure \ref{fig:charge_state_time} c) and required time (figure \ref{fig:charge_state_time} d) to initialise the state. 
We find that for the fidelity $99\%$, our method of estimation fidelity favours for the short time and high laser power, while the method accounting on initial state would be giving slightly different time, and underestimate the fidelity.

\subsection*{Nuclear spin readout at room temperature}
\begin{figure}[ht]
\begin{center}
\includegraphics[width=\columnwidth]{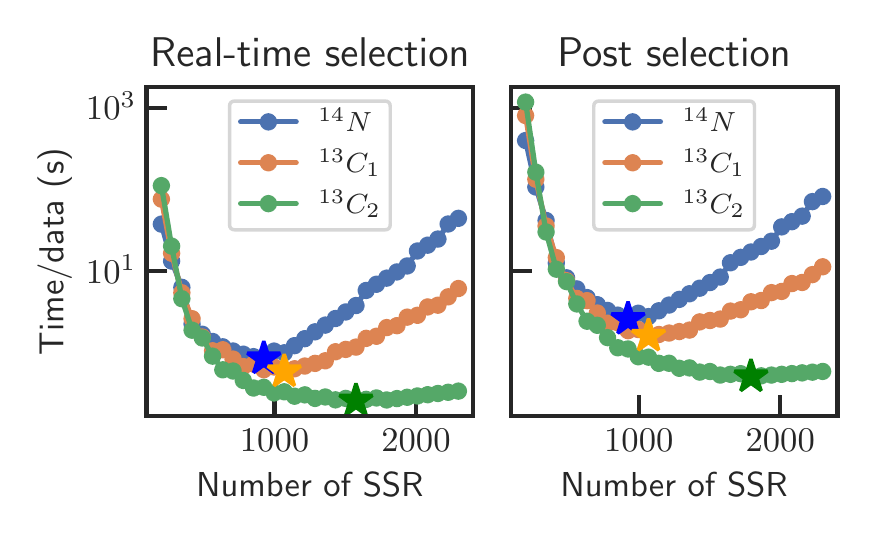}
\caption{The optimal number of repetitive readouts in the single-shot readout for the preparation of nuclear spins  a) The case of dynamical preparation via the feedforward and real-time control, b) The case of postselection.}
\label{fig:nuc_spin_time}
\end{center}
\end{figure}
We apply our approach for the strongly coupled nuclear spins near the NV centre at room temperature used as qubits. 
We consider two scenarios. 
In the first case, the initialisation is done by measurement for the qubit, and then the sequence, e.g., for sensing, is used. 
Second, the initialisation is done until $success$ (on demand), followed by the execution of the main sequence (sensing or quantum algorithm). 
We extract the decay rates for nuclear spins under the readout using the autocorrelation method of the time traces discussed in the SI.
Similar to the case of the charge state, we optimise the number of repetitive readouts to reach the target initialisation fidelity to perform a single successful measurement in the shortest time.   
By varying the number of repetitions of the CNOT gate with a green laser pulse \cite{NeumannSSR} and adjusting the threshold to reach the targeted fidelity of 99\%, we analyse the required number of attempts and the preparation time for each nuclear spin in the register expressed in the figure \ref{fig:nuc_spin_time}.
We find that in the first case of postselection, the optimal point is 2.6, 1.6, and 0.5 s for nuclei $^{14}N$, $^{13}C_1$, $^{13}C_2$, while for the case of on-demand preparation the average preparation time is shorter by 2-3 times 0.85, 0.59, 0.26 s correspondingly. 
The required time to generate one data point with 10 $\mu s$ microwave time, 50 $\mu s$ RF time, 10 ms average sequence time, and 10 ms readout time. 
Moreover, the optimum parameters for nuclear spins with the dynamical and postselection methods differ. We notice that the dynamical real-time on-demand preparation method requires a smaller number of SSR repetitions, indicating that the time cost of a single attempt is lower.



\section{Conclusion}
\label{sec:discussion}
Not only counting the number of photons but also considering photons' arrival time and their correlations will potentially provide additional information, which leads to better initialisation fidelities as was already shown for the readout \cite{Gambetta2007vx,DAnjou2014tz, DAnjou2016up,DAnjou2017us}. 
As opposed to the readout of the initial state, for initialisation, the photons that arrive later carry more information about the final state.
The exponentially growing linear and nonlinear methods of inferring the final state could improve the fidelity and could be studied.  
In conclusion, we formulated the method for accurately estimating and optimising the fidelity of the initialisation of the system state by finite demolition measurement.  
We considered three cases.
We find that parameters for initialisation are different from the readout when optimised for the required success time and should be optimised separately.  
We believe that our treatment is also applicable and interesting to other systems like dopants in SiC or rare earth ions \cite{Kindem_2020,Raha_2020,Anderson_2022}.

\section*{Acknowledgments}
The authors are thankful to Nikita Ratanov for fruitful discussions. M.Z. thanks Max Planck School of Photonics for financial support.
We acknowledge financial support by the European Union’s Horizon 2020 research and innovation programme via the project Quantum Internet Alliance (QIA, GA no. 820445),
Federal Ministry of Education and Research (BMBF) project MiLiQuant and Quamapolis, Spinning, QRx, the DFG (FOR 2724), the Land Baden-Württemberg via the project QC4BW,
The Max Planck Society, and the Volkswagentiftung.

\appendix
\section{Derivation of the Probabilities distribution for photon counts}
\label{app:theory}
We consider a case of a two-level system (TLS) with states $\ket{0}$ and $\ket{1}$. 
Under the asymmetric stationary decay with rates $\gamma_1, \gamma_2$ between the two states 0 and 1, respectively system performs the sequence of transitions (\textit{jumps}) events which form a point Poisson process. 
The time intervals $\{\tau_i\}$ and $\{t_i\}$ spend in state 0, and 1 between the switches are then random variables which have exponential distributions: $\tau \sim \mathrm{Exp}(\gamma_1),\, t \sim \mathrm{Exp}(\gamma_2) $. 
We recall the following known properties related to the exponential distributions.
 \paragraph{Lemma 1}: The sum of $n$ exponentially distributed random variables $x_i$ with rate $\gamma$: $X_n = \sum_i^n x_i$ is a random variable. It has Erlang distribution with probability density function 
 \begin{equation}
 p(X_n = x) = \gamma \frac{e^{-\gamma x}(\gamma x)^{n-1}}{(n-1)!}
 \end{equation}
 Additionally, we introduce a random variable $N_x$ \cite{metis}, which is defined as $N_x = \min (n | \sum_i^n x_i \ge x)$  and represents the minimum number of elements from a given set of random variable sample, which sum exceeds $x$. 
 \paragraph{Lemma 2}: Variable $N_x-1$ has Poisson distribution, and $N_x$ has a probability density function as follows
 \begin{equation}
 p(N_x = n) = \frac{e^{-\gamma x}(\gamma x)^{n-1}}{(n-1)!}.
 \end{equation}

Similar to work \cite{Shields2015}, we consider cases of odd and even numbers of switching events separately. 
We introduce $n$ as the number of intervals spent in state $\ket{0}$.
We can now estimate the probability that the system spends time $\tau$ in state $\ket{0}$ during the measurement time $T$. 
When having an odd number of switching, and starting from state $\ket{0}$ the intervals between switches are sets of random variables $\{\tau_i\}\sim \mathrm{Exp(\gamma_1)}$ and $\{t_i\}\sim \mathrm{Exp(\gamma_2)}$ with rates $\gamma_{1,2}$. 
The Probability that the system spends total time $\tau$ in state $\ket{0}$ is a sum of products of the Erlang-n distribution that sum of n variables equals $\tau$ with the probability that $n$ intervals occur, which is the probability that the residual time $t=T-x$ is exceeded in $n$ increments of a process $\{t_i\}$, hence:
\begin{equation}
\begin{split}
&p(\tau \cap odd |\ket{0}) = \sum_{n=1}^\infty P\left(\tau_n = \tau\right) P\left(N_{t=T-\tau} = n\right) \\
& = \gamma_1 e^{(\gamma_2 - \gamma_1 ) \tau -\gamma_2 T} \sum_{n=1}^\infty \frac{(\gamma_1 \tau \gamma_2 (T-\tau))^{n-1}}{(n-1)!^2} \\ 
& = \gamma_1 e^{(\gamma_2 - \gamma_1 ) \tau -\gamma_2 T} I_0 \left(2\sqrt{\gamma_1 \gamma_2 \tau (T-\tau)}\right), 
\end{split}
\end{equation} 
where $I_0(z)=\sum_{i=0}^\infty \frac{(z/2)^{2n}}{n!^2}$ is modified Bessel function of the first kind of $0-th$ order.
For the case of an even number of switches, we have to take the opposite consideration. The total interval $t = T-\tau$ has a fixed length, and the length $\tau$ has to be exceeded because the system could stay in the final state after time $T$. Hence the probability is sum over $n$ of probability that process $\{\tau_i\}$ exceeds $\tau$ in $n$ increments (steps), times the probability that $n-1$ intervals $\{t_i\}$ sum to $T-x$, hence:

\begin{equation}
\begin{split}
&p(\tau \cap even|\ket{0}) = \sum_{n=2}^\infty P \left( t_{n-1}=T-\tau \right) P\left(N_\tau = n \right) \\
& = \gamma_2 e^{(\gamma_2 - \gamma_1) \tau -\gamma_2 T} \sum_{n=2}^\infty \frac{(\gamma_1 \tau)^{n-1} (\gamma_2 (T-\tau))^{n-2}}{(n-1)!(n-2)!} \\ 
&= e^{(\gamma_2 - \gamma_1 ) \tau -\gamma_2 T} \cdot \sqrt{\frac{\gamma_1 \gamma_2 \tau}{T-\tau}} I_1 \left(2\sqrt{\gamma_1 \gamma_2 \tau (T-\tau)}\right), 
\end{split}
\end{equation}
where $I_1(z)=\sum_{i=0}^\infty \frac{(z/2)^{2n+1}}{n! n+1!}$ is the modified Bessel function of the first kind of $1-st$ order. Additionally, we consider the case where no switches happen, which simply reads: 
\begin{equation}
p(\tau=T | switchless, \ket{0}) = e^{-\gamma_1 T}
\end{equation}
The probability conditioned on initial state $\ket{1}$ could be obtained by substituting $\gamma_1 \leftrightarrow \gamma_2$ and $\tau\leftrightarrow T-\tau$.
Using the property of conditional probability, one obtains:
\begin{equation}
p(\tau \cap A | \ket{0}) + p(\tau \cap A^c | \ket{0}) = p(\tau |\ket{0}),
\end{equation} 
Where $A = even = odd^c$, and c denotes complementarity, we conclude the derivation of the distribution of the time spent by the system in the state $\ket{0}$ conditioned on the initial state: 
\begin{widetext}
\begin{equation}
p(\tau | \ket{0}) = e^{(\gamma_2-\gamma_1 ) \tau - \gamma_2 T} \left(\gamma_1 I_0\left(2\sqrt{\gamma_1\gamma_2\tau(T-\tau)}\right)+\sqrt{\frac{\gamma_1 \gamma_2 \tau}{T-\tau}} I_1\left(2\sqrt{\gamma_1 \gamma_2 \tau \left(T-\tau\right)}\right)\right)+e^{-\gamma_1 T}\delta(\tau - T)
\label{eq:prob_0}
\end{equation}
\begin{equation}
p(\tau | \ket{1}) = e^{\left(\gamma_1-\gamma_2\right) \left(T-\tau\right) - \gamma_1 T} \left(\gamma_2 I_0\left(2\sqrt{\gamma_1\gamma_2\tau\left(T-\tau\right)}\right)+\sqrt{\frac{\gamma_1 \gamma_2 \left(T-\tau\right)}{\tau}} I_1\left(2\sqrt{\gamma_1 \gamma_2 \tau \left(T-\tau\right)}\right)\right)+e^{-\gamma_2 T}\delta\left(\tau\right)
\label{eq:prob_1}
\end{equation}
\end{widetext}
\subsection*{Photon counting statistics conditioned on initial state}
Assuming emitted photons from the system arrive on the photodetector at random times with constant rate $\lambda_1$ and $\lambda_2$ conditioned on the system state. 
The number of photon counts is a random variable
\begin{equation}
\lambda \sim \mathrm{Poisson\left(\lambda_1 \tau + \lambda_2 (T-\tau)\right)}
\label{eq:pois}
\end{equation}
, where $T$ is the total counting time, and $\tau$ is the total time spent in state $\ket{0}$.
Using the expressions for probability density for $\tau$ eq. \ref{eq:prob_0}, eq.\ref{eq:prob_1}, combining with eq.\ref{eq:pois} and integrating $\tau$ over the interval $\tau \in \left[0, T\right]$ we obtain expression for the photon counting statistics similar to \cite{Shields2015}. 

\begin{widetext}
\begin{equation}
\begin{split}
p\left(\lambda | \ket{1}\right) =&\int_0^T  \Bigg\{ e^{\left(\gamma_1-\gamma_2 \right) \left(T-\tau\right) - \gamma_1 T} \left(\gamma_2 I_0\left(2\sqrt{\gamma_1\gamma_2\tau\left(T-\tau\right)}\right)+\sqrt{\frac{\gamma_1 \gamma_2 \left(T-\tau\right)}{\tau}} I_1\left(2\sqrt{\gamma_1 \gamma_2 \tau \left(T-\tau\right)}\right)\right) \times \\ 
& \mathrm{Poisson}\left(\lambda, \lambda_1 \tau + \lambda_2 (T-\tau)\right) d\tau  \Bigg\}+ e^{-\gamma_2 T} \mathrm{Poisson}\left(\lambda, \lambda_2 T\right)
\label{eq:prob_n}
\end{split}
\end{equation}
\begin{equation}
\begin{split}
p\left(\lambda | \ket{0}\right) =&\int_0^T  \Bigg\{ e^{\left(\gamma_2-\gamma_1 \right) \tau - \gamma_2 T} \left(\gamma_1 I_0\left(2\sqrt{\gamma_1\gamma_2\tau\left(T-\tau\right)}\right)+\sqrt{\frac{\gamma_1 \gamma_2 \tau}{T-\tau}} I_1\left(2\sqrt{\gamma_1 \gamma_2 \tau \left(T-\tau\right)}\right)\right) \times \\ 
& \mathrm{Poisson}\left(\lambda, \lambda_1 \tau + \lambda_2 (T-\tau)\right) d\tau  \Bigg\}+ e^{-\gamma_1 T} \mathrm{Poisson}\left(\lambda, \lambda_1 T\right)
\label{eq:prob_n}
\end{split}
\end{equation}
\end{widetext}
\bibliography{references}

\end{document}